\documentclass[%
 reprint,
superscriptaddress,
 amsmath,amssymb,
 aps,
 twocolumn,
]{revtex4}

\usepackage{graphicx}
\usepackage{dcolumn}
\usepackage{bm}

\usepackage[dvipsnames]{xcolor}
\usepackage{bm}
\usepackage{comment}

\begin{document}

\title{Single-Molecule Structure and Topology of Kinetoplast DNA Networks}

\author{Pinyao He}
\affiliation{Department of Bionanoscience, Kavli Institute of Nanoscience, Delft University of Technology, 2629 HZ Delft, Netherland}
\affiliation{Jiangsu Key Laboratory for Design and Manufacture of Micro-Nano Biomedical Instruments, School of Mechanical Engineering, Southeast University, Nanjing 211189, China}
\author{Allard J. Katan}
\affiliation{Department of Bionanoscience, Kavli Institute of Nanoscience, Delft University of Technology, 2629 HZ Delft, Netherland}
\author{Luca Tubiana}
\affiliation{Physics Department, University of Trento, via Sommarive, 14 I-38123 Trento, Italy}
\affiliation{INFN-TIFPA, Trento Institute for Fundamental Physics and Applications, I-38123 Trento, Italy}
\affiliation{Faculty  of  Physics,  University  of  Vienna,  Boltzmanngasse  5,  1090  Vienna,  Austria}
\author{Cees Dekker}
\affiliation{Department of Bionanoscience, Kavli Institute of Nanoscience, Delft University of Technology, 2629 HZ Delft, Netherland}
\author{Davide Michieletto}
\thanks{corresponding author, davide.michieletto@ed.ac.uk}
\affiliation{School of Physics and Astronomy, University of Edinburgh, Peter Guthrie Tait Road, Edinburgh, EH9 3FD, UK}
\affiliation{MRC Human Genetics Unit, Institute of Genetics and Cancer, University of Edinburgh, Edinburgh EH4 2XU, UK}

\begin{abstract}
\textbf{The Kinetoplast DNA (kDNA) is a two-dimensional Olympic-ring-like network of mutually linked 2.5 kb-long DNA minicircles found in certain parasites called Trypanosomes. Understanding the self-assembly and replication of this structure are not only major open questions in biology but can also inform the design of synthetic topological materials. Here we report the first high-resolution, single-molecule study of kDNA network topology using AFM and steered molecular dynamics simulations. We map out the DNA density within the network and the distribution of linking number and valence of the minicircles. We also characterise the DNA hubs that surround the network and show that they cause a buckling transition akin to that of a 2D elastic thermal sheet in the bulk. Intriguingly, we observe a broad distribution of density and valence of the minicircles, indicating heterogeneous network structure and individualism of different kDNA structures. Our findings explain outstanding questions in the field and offer single-molecule insights into the properties of a unique topological material. 
}
\end{abstract}

\maketitle

\section*{Introduction}
The Kinetoplast DNA (kDNA) is one of the most fascinating naturally occurring genomes~\cite{Simpson1976,Laurent1970,Shlomai1983,Perez-Morga1993,Shlomai1994,Morris2001,Lukes2002}. It is formed in the mitochondrion of unicelluar parasites of the class Kinetoplastida and it is composed by an interlinked two-dimensional network of small DNA circles, or ``mini-circles'' and larger DNA rings called ``maxi-circles''. Maxicircles contain the genetic information for the synthesis of mitochondrial proteins, while the minicircles display somewhat redundant genetic information and are mainly necessary to perform extensive RNA editing on the maxicircles mRNA~\cite{Simpson1967}. The precise composition of the network depends on the organism; for instance, \emph{Crithidia fasciculata} (\emph{C. fasciculata}) kDNA is contained within a $1\mu$m $\times$ $0.5\mu$m disk-shaped organelle and made of about 5000 mini-circles (2.5 kb, or 850 nm long) and 30 maxicircles (about 30 kb, or 10 $\mu$m long). The mechanisms through which kDNA self-assembles and replicates are poorly understood~\cite{Perez-Morga1993,Liu2005,Klingbeil2001a}. 

The evolutionary benefit of a linked mitochondrial genome remains a major open question in Trypanosome biology~\cite{Liu2005,Schnaufer2010}. It has been speculated that the interconnected structure of linked rings provides genomic stability and a means to mechanically preserve genetic material, i.e. to avoid losing minicircles during cell division~\cite{Liu2005}. A common feature of kDNA is that it is found in the basal body, near the parasite flagellum. For this reason, it has also been speculated that the linkedness of the network may serve to provide mechanical stability to the organelle~\cite{Morris2001}.  Taken outside the parasite, kDNA expands to assume a ``shower-cap'' buckled shape about 5$\mu$m in size~\cite{Klotz2020,Soh2020,Soh2021}. Once adsorbed onto a surface for Electron Microscopy (EM) or atomic force microscopy (AFM), kDNA stretches to an oval shape 8$\mu$m $\times$ 10$\mu$m in size and displays a thick border which is characterised by rosettes and brighter nodes~\cite{Fairlamb1978,Cavalcanti2011,Yaffe2021,barker1980ultrastructure}. 

In 1995, Cozzarelli and coauthors designed an elegant, albeit indirect, bulk method based on gel electrophoresis of digestion products to show that \emph{C. fasciculata} kDNA topology is compatible with a two-dimensional hexagonal network where each ring is linked to other three minicircles, on average~\cite{Chen1995,Chen1995a}. 
These results have also been recently independently confirmed using the same bulk method~\cite{Ibrahim2019}. 
In spite of this, recent microscopy experiments indicate that kDNAs assume highly heterogeneous shapes, suggesting a broad spectrum of topologies~\cite{Klotz2020}. 
 
\begin{figure*}[th!]
    \centering
    \includegraphics[width=1\textwidth]{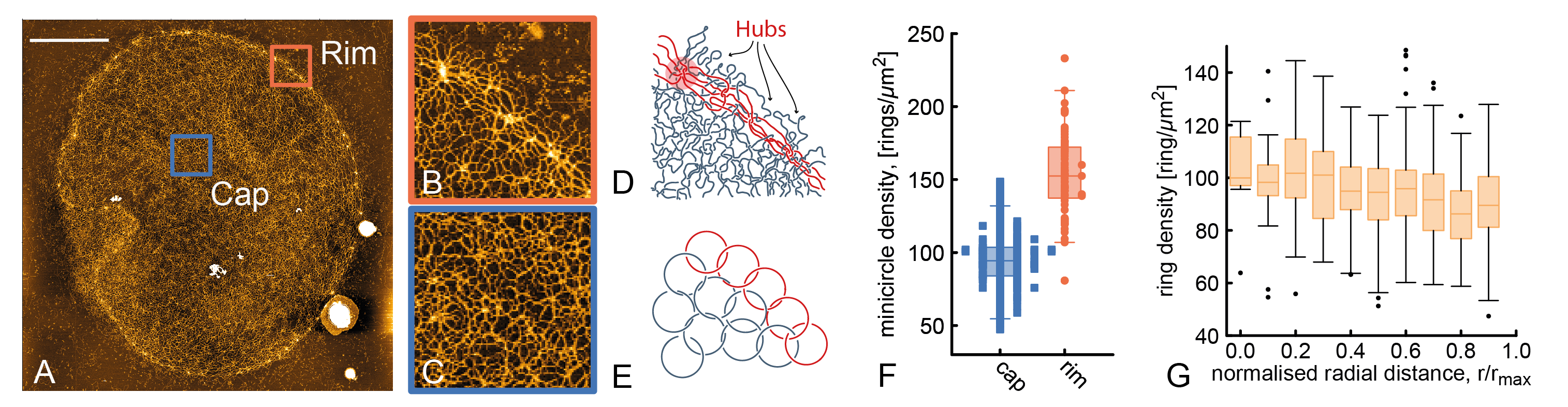}
    \caption{\textbf{A} AFM images of kDNA from {\it C. fasciculata}. The scale bar is 2 $\mu$m and the yellow color scale ranges from 0 (black) to 3.5 nm height (white). \textbf{B-C} Zoom ins of the rim and cap respectively. \textbf{D} Sketch of the network of minicircles from \textbf{B} where we color-coded the minicircles forming the rim in red. \textbf{E} Simplified sketch where we schematically show that the network is formed by linked rings. \textbf{F} Boxplot of minicircle density in the cap and the rim of the network (obtained from selected regions and averaged across 3 kDNA networks). The rim density is computed by taking a circular region of radius $r=100$ nm (about the size of a minicircle, see below) centred at the hubs. \textbf{D} Boxplots of the density of minicircles as a function of distance from the centre. } 
    \label{fig:panel_density}
\end{figure*}

Alongside experiments, computational and theoretical work have provided evidence that this type of linked network may be formed as a result of a percolation transition~\cite{Diao2012,Michieletto2014a,Ibrahim2019}. Beyond the percolation transition, overlapping rings form a system-spanning network of interlocks. At the onset percolation the mean valence $v$, i.e. the number of rings that are linked to any one ring on average, was found to be 3~\cite{Michieletto2014a}, in agreement with gel electrophoresis experiments~\cite{Chen1995}. 

Arguably, the minicircles acquire their valence {\em in vivo}, where the kDNA is under large confinement. Given a minicircles number density $\rho \simeq 5000 \textrm{rings}/(\pi (0.5 \mu\textrm{m})^3) \simeq 12700 \, \textrm{rings}/\mu$m$^{3}$ and a radius of gyration of a minicircles $R_g \simeq l_p \sqrt{L/12 l_p} \simeq 60$ nm, one would expect a number of overlaps per minicircle $P = 4 \rho /3 \pi R_g^3 \simeq 11.5$. Even if only half of the overlapping minicircles became linked to each other via Topoisomerase-mediated strand-crossing~\cite{Shlomai1983} or the linking effectively occurred in 2D due to stacking and alignment~\cite{Diao2012,Silver1986}, this overlapping number would still yield a valence much larger than $v=3$ estimated by Cozzarelli~\cite{Chen1995}.
These arguments suggest that the kDNA cannot be thought of a gas of freely crossable rings, and instead regulates its topology via, e.g., packaging proteins such as KAP~\cite{Yaffe2021} or by tuning the activity of Topoisomerases. 


All of the quantitative evidence on kDNA network topology comes from indirect, bulk measurements~\cite{Chen1995,Chen1995a,Ibrahim2019} and recent microscopy work suggests that different kDNA networks have very different shape and behaviours, suggesting heterogeneity in the self-assembly of this fascinating structure~\cite{Klotz2020}. To shed more light into this, here we study {\em C. fasciculata} kDNA networks using single-molecule techniques (AFM) and Molecular Dynamics Simulations. More specifically, we first quantitatively map the density of minicircles in the network as a function of their position and quantify the network structure by measuring its porosity. We also identify the characteristic rosettes at the rim of the network as originating from the localisation of essential crossings. Imposing a constraint on the size of the rim, we show that the network undergoes a buckling transition that explains recent {\it in vitro} observations. Then, we employ steered molecular dynamics simulations to reconstruct the topology of the network at the single-molecule level. We thus obtain the full distribution of the valence in the network: we find it to be compatible with a valence 3 but at the same time displays a broad distribution suggesting heterogeneity in the network topology and across networks. Notably, our findings are not compatible with a perfect hexagonal network thus refusing the classical model by Cozzarelli~\cite{Chen1995}. Finally, we discuss our findings in light of the work done on sub-isostatic floppy networks~\cite{Maxwell1864} and 2D elastic thermal sheets~\cite{Shankar2021} and predict that the kDNA should display a Young modulus much lower than that of common 2D materials such as lipid membranes.  

\section*{Results}
\subsection*{The density of minicircles is not uniform}

\begin{figure}[t!]
    \centering
    \includegraphics[width=0.5\textwidth]{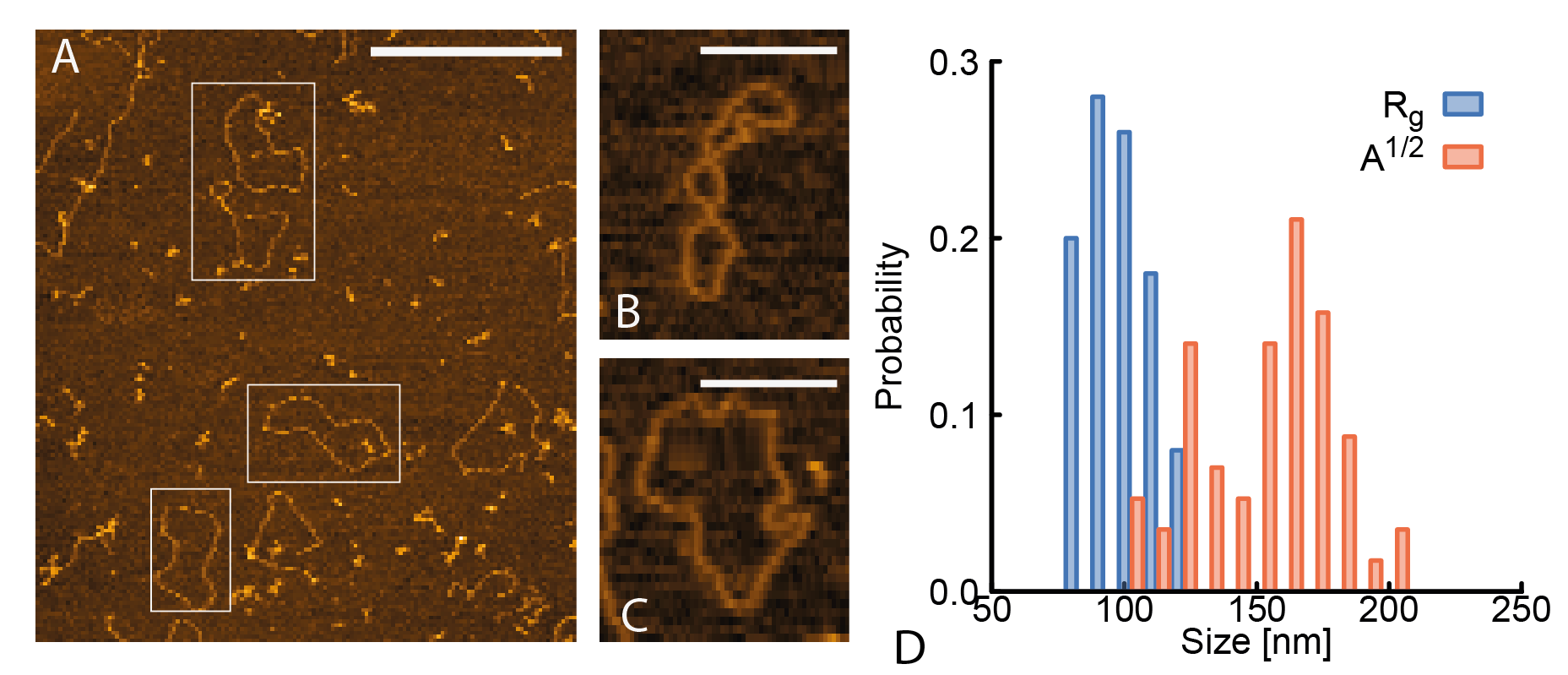}
    \vspace{-0.5 cm}
    \caption{\textbf{A} Examples of single rings used to compute the perimeter and radius of gyration of single minicircles. Scale bar is 500 nm. \textbf{B-C} Zoom ins of minicircles with supercoiled-like (\textbf{B}) and open (\textbf{C}) conformations. Scale bars are 250 nm. \textbf{D} Size of the minicircles, computed as the radius of gyration of the AFM traces, $R_g$, and the square root of the area, $A^{1/2}$. We find a mean $R_g = 101.3 \pm 10.8$ nm, compatible with Ref.~\cite{Wirtz2009} reporting $R_g = 109$ nm for 2.6 kb-long plasmids, and mean $A^{1/2} = 154.4 \pm 22.5$ nm. }
    \vspace{-0.5 cm}
    \label{fig:panel_single_rings}
\end{figure}

We perform dry, high-resolution AFM on \emph{C. fasciculata}  kDNA (TopoGen). A representative image, zoom ins and sketches are shown in Fig.~\ref{fig:panel_density}B-E. We first noticed that the networks displayed fluctuations in the density of minicircles (bright/dim areas within the kDNA ``cap''). At the edge of the networks, we noticed bright and regularly spaced nodes along its edge, as previously reported~\cite{barker1980ultrastructure,Cavalcanti2011,Yaffe2021} (see Fig.~\ref{fig:panel_density}B-E). 
To quantify the density of minicircles in different regions of the kDNA we first measured the volume of isolated minicircles outside the kDNA network (see also Fig.~\ref{fig:panel_single_rings}). These provided an internal control in our experiments, as minicircles outside the kDNA network are subjected to the same experimental artefacts (e.g. sample dehydration) than the ones within the network. In turn, we obtained an average volume for the single mini-circles $I_{mc}$ which we used to normalise the volume found in regions within the kDNA. We then randomly sampled selected regions within the cap of the network and normalised their total volume $I(\bm{r})$ by $I_{mc}$. The quantity $\rho(\bm{r}) \equiv I(\bm{r})/(I_{mc} A)$, where $A$ is the area of the sampled region, is the number density of minicircles at location $\bm{r}$ in the image.

By averaging over 3 independent kDNA networks (see SI for raw images), we find that there are $\rho = 94 \pm 17$ rings/$\mu$m$^2$ in the cap (see Fig.~\ref{fig:panel_density}F). Given that the mean short and long axes of our kDNAs are $l = 7.8$ $\mu$m and $L=9.1$ $\mu$m, respectively, we find a corresponding total number of minicircles $N_{tot} = \rho \pi l L  = 5296$. Considering the limits of the pixel resolution and the assumptions made for the conversion of signal intensity to DNA mass, this number is in excellent agreement with that reported in the literature, i.e. $5000$~\cite{Chen1995} for \emph{C. fasciculata} kDNA. It should be highlighted that we could have arrived at a similar value of $\rho$ by simply assuming that the network is formed by $5000$ rings uniformly distributed in the network, yet using our method we have (i) verified independently that the network has around $5000$ rings and (ii) developed a way to measure ring density as a function of position $\bm{r}$ in the kDNA. 
By applying the same method to the hubs along the rim of the kDNA, which we define as the region within one one minicircle size ($r \simeq 100$ nm, see below) from the centre of the brightest nodes (see Fig.~\ref{fig:panel_density}B), we find that the average ring density is significantly larger, with mean $\rho_{\rm rim} = 153.1 \pm 27.0$ rings/$\mu$m$^2$. 
 
We then asked if there was a dependence of minicircle density as a function of position in the network. We sampled about 200 small regions in 3 different kDNA networks and computed $\rho(\bm{r})$ as above. We then plotted this as a function of the radial distance $r = |\bm{r}-\bm{r}_c|$ from the centre of the network. We discovered that the density displays a smooth decrease by $\sim$ 13\% from the centre to the periphery (Fig.~\ref{fig:panel_density}D). Since a uniformly filled disk that is stretched isotropically will still display a uniform mass distribution we argue that the observed density gradient is a feature of the network rather than an artefact of the imaging method. We hypothesise that this density gradient may be locked in at the end of replication -- which occurs at the antipodal points positioned outside the kDNA in this Trypanosome species -- as the mini-circles can no longer unlink from the neighbours and redistribute within the network due to the absence of type 2 Topoisomerase.

The gradient in minicircle density suggests that the topology of the kDNA may not be uniform as minicircles in the middle of the cap may be more connected than the ones at the periphery (excluding the rim). The density gradient and the difference between DNA density in cap and rim has not been reported nor quantified before and we argue that these are potentially important to account for in future models of kDNA self-assembly~\cite{Chen1995,Diao2012,Michieletto2014a,Polson2021,Ibrahim2019}.

\subsection*{Estimating the valence of minicircles}
Based on our measurement of minicircle density within the cap, we now estimate the valence $v$ of the minicircles, i.e. the number of minicircles that are linked to any one minicircle. To do this we first compute the minicircles average size by tracing the contour of DNA rings found outside the network (in Fig.~\ref{fig:panel_single_rings}A, we show examples of minicircles used for this analysis). We find that isolated minicircles have a mean contour length of $L_c = 791 \pm 66$ nm and a mean radius of gyration $R_g = 101.3 \pm 10.8$ nm, which is compatible with the size measured for DNA plasmids of similar length absorbed in 2D~\cite{Witz2008}. Since we observed heterogeneous conformations displaying plectonemic-like writhe (Fig.~\ref{fig:panel_single_rings}B), we also computed the area of the minicircles and noticed that it displays a broader distribution, compatible with the presence of writhing and open minicircles in the AFM images (Fig.~\ref{fig:panel_single_rings}B-D). 
From the number density of minicircles per unit area $\rho$ and their average size $R_g$, we estimate that the number of overlapping minicircles in the flattened kDNA is $P \simeq \rho \pi R_g^2 \simeq 3$ (valid for isotropic and randomly shaped minicircles). This number is about 4-fold smaller than the number of overlaps expected {\it in vivo} (where we recall that the kDNA is contained within a disk 1 $\mu$m diameter and 0.5 $\mu$m height) but is compatible with the average valence $v \simeq 3$ measured by Cozzarelli~\cite{Chen1995}. 

At the rim, we may use an effectively larger minicircle density, yielding $P \simeq \rho_{\rm rim} \pi R_g^2 \simeq 4.8$ in turn suggesting a larger valence of the minicircles at the hubs. However, we note that the minicircles at the rim are stretched, in turn increasing their $R_g$ and potentially their real valence. In the next section we shall characterise the minicircles at the rim in more detail.

Finally, we note that the $R_g$ we measured from the 2D absorbed minicircles is typically larger than the $R_g$ they would assume in bulk~\cite{Rivetti1996}. In the extreme case that they assume the shape of ideal loops, we recall that we would expect $R_g \simeq 60$ nm. In turn, we would expect a valence of about 1 for a ring density $\rho \simeq 95$ \mbox{$\mu$m$^2$}. On the other hand, we know that {\em in vivo} the kDNA is packaged at much larger density which will therefore increase the maximum valence that each minicircle can reach up. Indeed, if every minicircle were linked to every overlapping neighbour we would expect $v \gtrsim 10$. In spite of this, one should bear in mind that DNA minicircles cannot link without the presence of (type 2) Topoisomerase; thus, its activity within or at the periphery of the network appears to be critical to regulate the catenation of the network~\cite{Liu2005,Perez-Morga1993b,Rauch1993}. 

We should also note that the density of minicircles we measured in the previous section, together with the size of the minicircles, is a intimately related to the inherent topology of the network cap. If the minicircles had a larger valence, we would inevitably expect a correspondigly larger DNA density.


\subsection*{The kDNA hubs are sites of essential crossings between linked minicircles}
\begin{figure}[t!]
    \centering
    \includegraphics[width=0.5\textwidth]{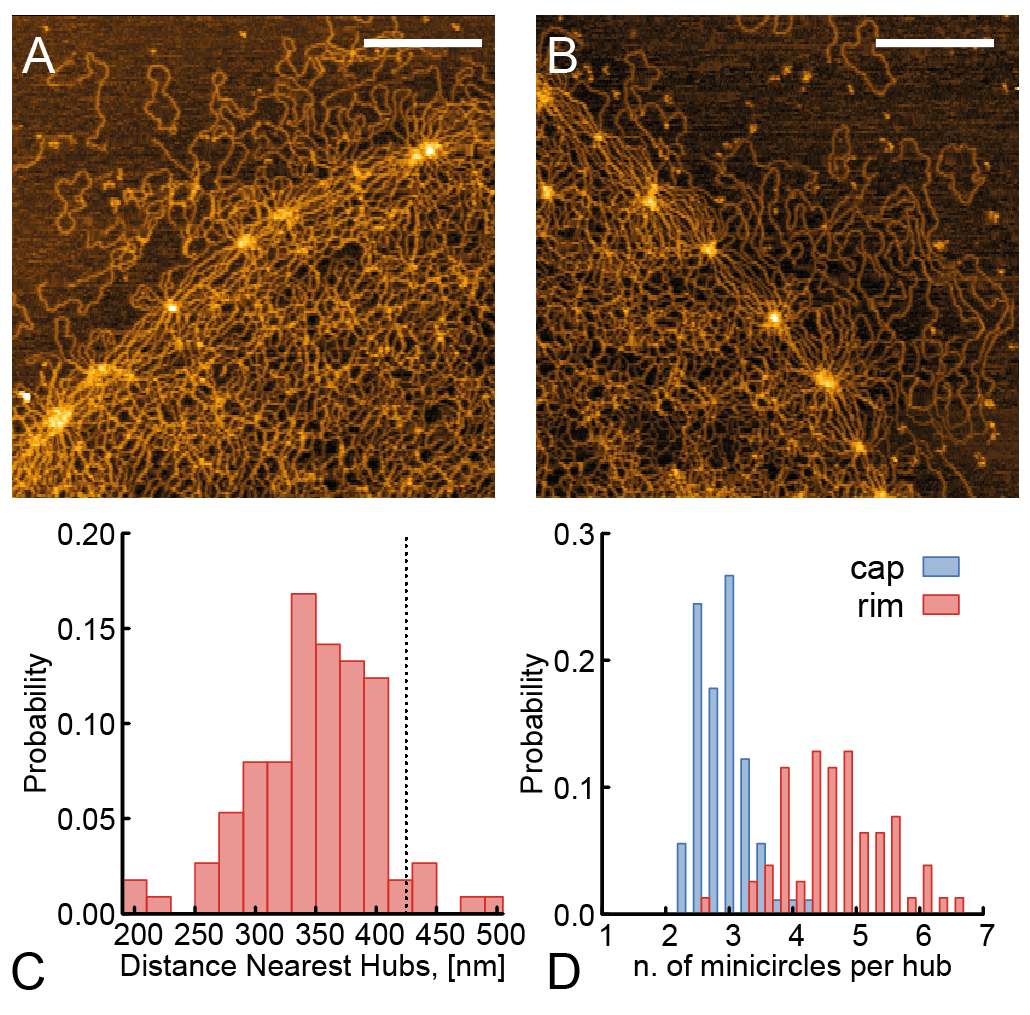}
    \vspace{-0.5 cm}
    \caption{\textbf{A-B} Zoomed in portions of kDNA showing hubs at the periphery. Scale bars are 500 nm. \textbf{C} Distribution of distances between nearest hubs. The dashed vertical line represents the diameter of a minicircle if pulled taut, i.e. $1/2 \times 2500$ bp $\times 0.34$ nm/bp $= 425$ nm. \textbf{D} Distribution of number of overlapping minicircles in the cap and in the hubs. }
    \vspace{-0.3 cm}
    \label{fig:panel_hubs}
\end{figure}

As mentioned above, a feature that stands out from the AFM images is the rim, formed by nodes (or hubs) connected by clear DNA tethers (Fig.~\ref{fig:panel_hubs}A-B). By zooming in these features one can appreciate that these nodes are formed when several minicircles come together into so-called ``rosettes''~\cite{barker1980ultrastructure} (Fig.~\ref{fig:panel_hubs}A-B). 

\begin{figure*}[t!]
    \centering
    \includegraphics[width=1\textwidth]{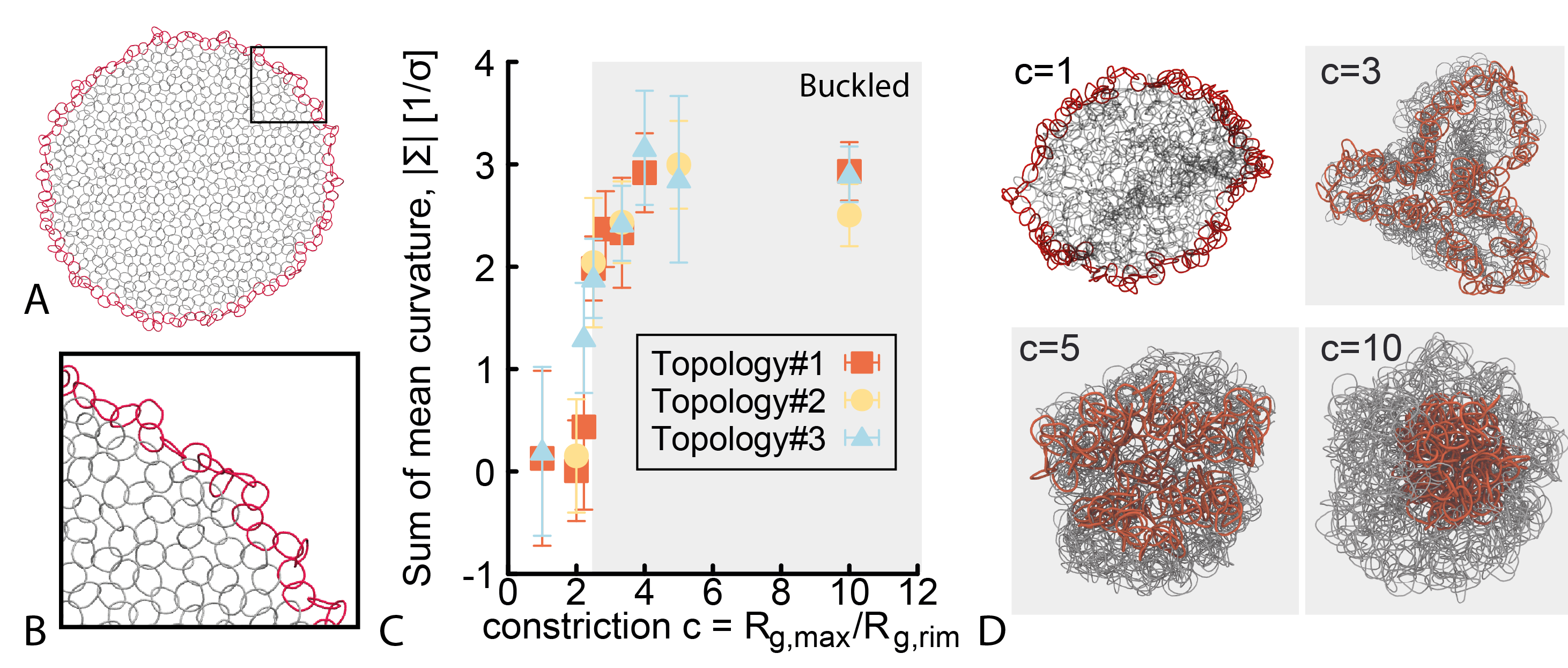}
    \caption{\textbf{A} An {\it in vitro} kDNA with hexagonal structure of rings. The border is highlighted in red. \textbf{B} Zoom in portion of \textbf{A} showing the hexagonal lattice linkages. See SM for details on how the network was built. \textbf{C} Absolute value of the sum of the mean curvature of the triangular mesh, $|\Sigma|$, calculated by joining the center of mass of the rings and plotted as a function of the constriction $c=R_{g,max}/R_{g,rim}$. In these simulations $R_{g,rim}$ is user-defined and steered using an harmonic potential. The plot shows an abrupt buckling transition. Different symbols correspond to different network topologies chosen by random linking of neighbouring rings while preserving the hexagonal structure. \textbf{D} Snapshots of the network at four different values of constriction $c$.}
    \label{fig:border_effect_sim}
\end{figure*}

The average distance between nearest nodes in the network is close to that of a minicircle pulled taut, i.e. $L_t= 0.5 \times 2500$ bp $\times$ $0.34$ nm/bp $ = 425$ nm (Fig.~\ref{fig:panel_hubs}C). Additionally, by directly measuring the density of strands in a circular region with radius $r = 100$ nm (equal to that of a minicircle in equilibrium outside the network) and centered at the nodes, we find that average number of overlapping minicircles per hub is $P_{\rm hub} = 4.7 \pm 0.8$, which should be compared with $P_{\rm cap} = 2.9 \pm 0.4$ found in the cap (Fig.~\ref{fig:panel_hubs}D). This may still be an underestimate, as the minicircles at the rim are stretched and their overlap number may thus be larger.  
Our images (see Fig.~\ref{fig:panel_hubs}A,B) also suggest that nearest nodes are directly connected by single minicircles, which are therefore redundantly linked. When minicircles are stretched due to the kDNA being absorbed, the essential crossings between minicircles become localised in hot-spots~\cite{Caraglio2017a}, forming the hubs. In the bulk, we expect the minicircles to relax and the hubs to disappear, although a rim with higher DNA density can still be visualised~\cite{Klotz2020}.

The reason why {\em C. fasciculata} kDNA displays a larger density of minicircles at the rim may be due to its replication mechanism, as newly replicated minicircles are added at the periphery from the antipodal points located just outside of a rotating kDNA network~\cite{Perez-Morga1993b,Chen1995a}. We also recall that {\it in vivo}, the kDNA is compressed in a disk of radius 0.5$\mu$m while it reaches $\sim$ 10 $\mu$m when fully adsorbed. This $\sim$20-fold compression effectively reduces the distance between nodes so that we expect the essential crossings to be within $1 R_g$ of each other. This redundancy in number of links is akin to that of replicating kDNA networks~\cite{Chen1995a}, and we thus argue that edge of the network could be made by newly replicated minicircles.

\subsection*{Simulations of kDNA with redundantly linked rim explain the buckling seen in bulk}

Our data suggests that the minicircles at the rim are both redundantly linked and stretched upon adsorption on the mica. In the bulk, these minicircles will tend to relax to their equilibrium diameter, i.e. reducing their size from $\simeq 350$ nm to $\simeq 170$ nm. This yields a $>$2-fold decrease in perimeter length as the essential crossings will become delocalised due to the minicircles' entropy in turn leading to a certain degree of overlap in between the minicircles~\cite{Caraglio2017a}. If we thus treat the perimeter of the kDNA as a poly-catenane (a polymer made of linked rings~\cite{Rauscher2018}) we can study the behaviour of a two-dimensional patch of linked rings under a varying constraint on its perimeter. In other words, we can study the behaviour of the kDNA under a variable degree of constraint on the length of its perimeter (due to minicircles relaxing to equilibrium conformations) by imposing a constraint on the size (radius of gyration) of the rim.

To do this we simulated a circular patch of an hexagonal (in line with Refs.~\cite{Chen1995,Polson2021}) network and constrained its border to have a radius of gyration $R_{g,rim}$ different from that assumed when the patch is completely planar, called $R_{g,max}$. We considered a system made of $n=604$ rings, each made of $m=60$ self-avoiding beads connected by FENE springs. For computational efficiency we considered semi-rigid rings, with persistence length $l_p = 120$ beads but we expect a similar result for flexible rings (see Fig~\ref{fig:border_effect_sim}A-B and SM). The network is constructed by placing the rings on the nodes of a circular patch of hexagonal lattice~\cite{hagberg2008exploring}, and then linking each ring with its neighbours choosing randomly between a $+1$ and a $-1$ Hopf link. We identify the border of the network as the set of rings having only two neighbours, or being directly linked to a ring having 2 neighbours only, and constrain its radius of gyration $R_g$ to a user-defined value $R_{g,rim}$ with an additional harmonic potential in the Hamiltonian 
\begin{equation}
V_{\rm constr}= K(R_g-R_{g,rim})^2 \, .    
\end{equation}
We then simulate the equilibrium behaviour of three different network topologies (distinguished by the sign of the linking numbers between neigbouring rings) for different values of $R_{g,rim}$ via Langevin dynamics in LAMMPS~\cite{Plimpton1995} (see SM for details). To characterize the equilibrium geometrical properties of the network we first map the hexagonal lattice of rings to a triangular mesh with edges connecting the center of mass of half of the rings and then compute the mean curvature as~\cite{fang2022topocut}  
\begin{equation}
    \Sigma = \sum_{i=1}^N \frac{1}{2}(K_{1,i} + K_{2,i})
\end{equation}
where $K_{1,i}$ and $K_{2,i}$ are the principal curvatures at facet $i$ (see SM for the details). The results, reported in Fig~\ref{fig:border_effect_sim}C, show that when the constraint at the perimeter is $R_{g,max}/R_{g,rim} > 2$, the equilibrium conformations display a buckled, ``shower-cap'' shape, as seen in experiments~\cite{Klotz2020} (see Fig.~\ref{fig:border_effect_sim}D). Interestingly, the absolute mean curvature $| \Sigma |$ increases abruptly from $0$ (saddle-like surface) to $\simeq 1 \sigma^{-1}$ (buckled) and eventually to $\simeq 3 \sigma^{-1}$ (shower-cap).

We recall that according to our measurements in the previous section, we expect the minicircles at the nodes to reduce their size from $\simeq 350$ nm to $\simeq 170$ nm when non-adsorbed to the mica. This implies a $>2$ fold shrinking in perimeter length (accounting for the fact that the essential crossings are localised when minicircles are stretched on the mica but partially delocalised when in bulk). Our simulations thus strongly suggest that the way minicircles are redundantly linked at the periphery is enough to induce the observed buckling.

\subsection*{Mesh size distribution}

Having quantified the distribution of minicircles in the network, we now quantify its mesh size.  Given that the minicircle density is around $\rho = 94$ rings/$\mu$m$^2$, the inter-minicircle separation is $\lambda = 1/\sqrt{\rho} = 103$ nm. In turn, we can estimate the mesh size as $\xi = |\lambda - 2 R_g| \simeq 100$ nm (recall that $R_g = 101.3$ nm, see Fig.~\ref{fig:panel_single_rings}). 

\begin{figure*}[t!]
    \centering
    \includegraphics[width=1\textwidth]{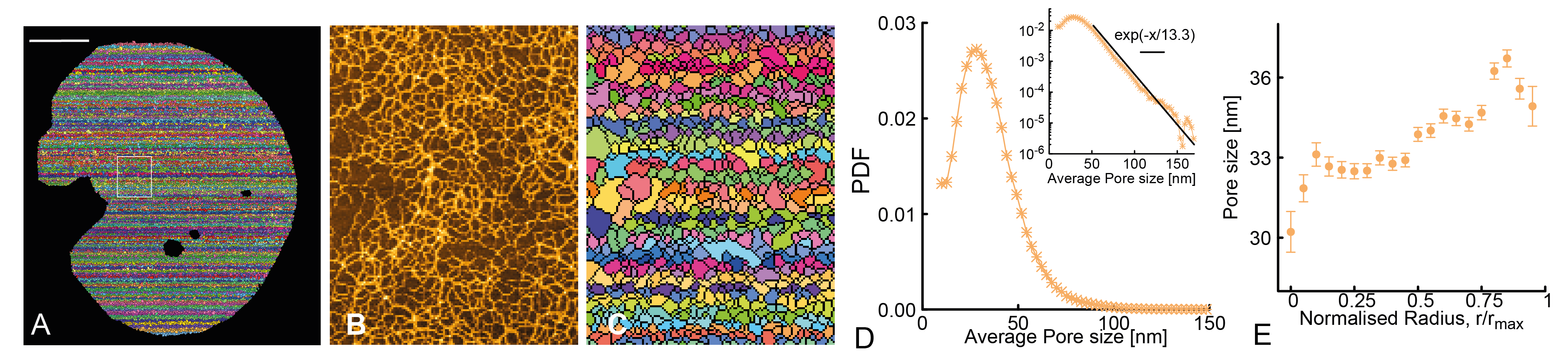}
    \caption{\textbf{A} Morphological segmentation of the AFM image show in Fig.~\ref{fig:panel_density}A. Scale bar is $1 \mu$m. \textbf{B} A zoomed in region showing side-by-side the AFM picture and in \textbf{C} the result of morphological segmentation. \textbf{D} Probability density function (PDF) of mesh size. Inset shows a log-linear plot of the same PDF, with an exponential decay reported as a guide for the eye. \textbf{E} Average pore size as a function of radial distance from the centre of the network.}
    \label{fig:poresize}
\end{figure*}

To quantify the mesh size more precisely, we employed morphological segmentation~\cite{Legland2016} to quantify the distribution of mesh sizes from our images. We first manually removed both imaging artefacts and the redundantly linked rim from within the region of interest (see Fig.~\ref{fig:poresize}A to be compared with Fig.~\ref{fig:panel_density}A). We then applied morphological segmentation to obtain a map of watershed basins, as shown in Fig.~\ref{fig:poresize}A-C. We then measure the area of each basin $a$ and estimate their size $\xi = \sqrt{a}$. We find that the values of pore sizes are broadly distributed and range between 10 and 200 nm with a peak (median) around $20$ nm and mean $\xi = 34.0 \pm 16.6$ nm. 

The distribution of mesh sizes appears to follow an exponential behaviour for values larger than $50$ nm (Fig.~\ref{fig:poresize}D). These values are small compared with the typical 100-500 nm of agarose gels and closer to those of DNA nanostar gels~\cite{Conrad2019}.

Interestingly, by computing the distance of all the basins from the centre of the network we observe that the average pore size increases towards the periphery (Fig.~\ref{fig:poresize}D). This is in line with our previous finding that the DNA density decreases towards the periphery. More specifically, we find pore sizes around 30 $\pm 0.7$ nm within the centre and around 35 $\pm 0.7$ nm near the periphery (where the redundantly rim was excluded). We note that this smaller than the crude calculation we made above, which is valid only for perfectly rigid minicircles. This is most likely due to the fact that the minicircles are writhing onto themselves thereby yielding smaller pore sizes overall.  

Finally, we note that in a lattice of rigid rings where every overlap is a link, one can map the number of pores (basins in the morphological segmentation) to the number of rings and their valence as $ N_{pores} > N_{rings} (1 + v/2)$. With flexible rings and non-connected overlaps, we expect more pores formed. We can therefore set an upper bound on the valence in the networks that we analyze,  $v < 2(N_{pores} - N_{rings})/N_{rings}$. From the morphological segmentation we find $N_{pores} = 19644 \pm 2404$ in turn implying $v < 6$. This large upper bound is likely due to overlaps which do not result in linking, yet still create mesh pores. Arguably, and contrary to common chemically crosslinked gels, the observed pores only mildly contribute to the network elasticity, as these overlaps can be easily removed by pulling the rings past each other.

\begin{figure*}[t!]
    \centering
    \includegraphics[width=1\textwidth]{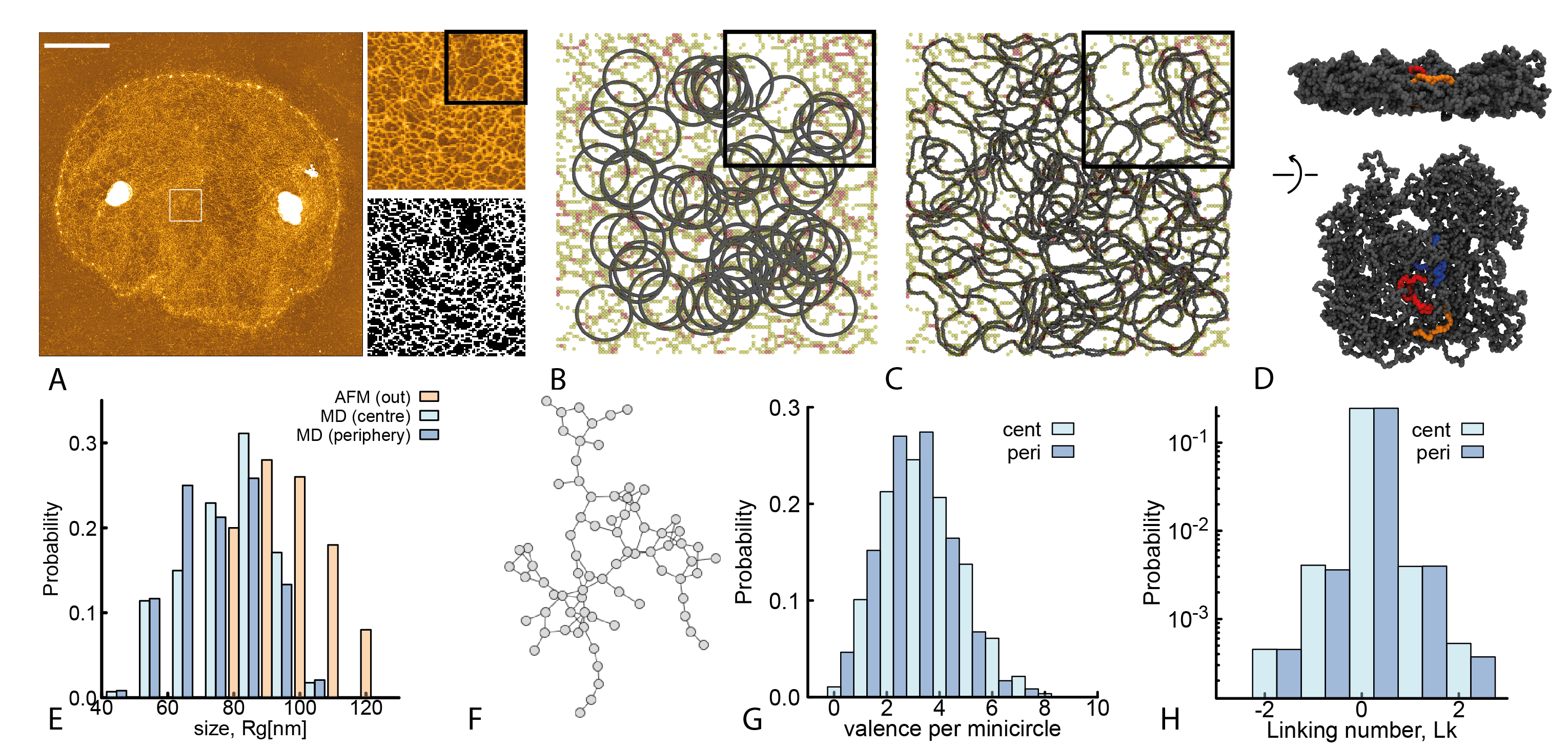}
    \caption{\textbf{A} Figure showing a kDNA image (scale bar is 2 $\mu m$) and a zoomed in 1 $\mu$m$^2$ region with the corresponding thresholded image (black and white). \textbf{B} Starting configuration of the AFM steered MD: 90 rings are placed randomly in a quasi two-dimensional simulation box. Phantom beads (yellow, orange and red in the figure) are static and act as Gaussian attracting basins. \textbf{C} At the end of the MD simulation, we obtain ensembles of conformations that capture the correct length and size of minicircles, display no ambiguous crossings with other rings and are compatible with the underlying AFM image. \textbf{D} Snapshot of the resulting network after having removed the slab confinement. \textbf{E} Distributions of $R_g$ of the minicircles from AFM (picked from outside the kDNA) and of the simulated ones (both from the centre and periphery). \textbf{F} Network representation of \textbf{D}, where each node is a minicircle and an edge between nodes represents that they are linked. Our simulated networks typically have all the nodes in one large connected component. \textbf{G} Distribution of valence, showing that the central minicirlces are on average more connected that the peripheral ones. \textbf{H} Distribution of linking number, showing that the most linked pairs are singly linked, and about 10\% of them are doubly linked.}
    \label{fig:MD}
\end{figure*}

\subsection*{AFM-steered simulations}
Despite the high-resolution AFM images, it is still challenging to identify single minicircles inside the cap of the network, and even less clear is to identify over/under-crossings between chains (see for example Fig.~\ref{fig:poresize}B). Because of this, it is impossible to unambiguously compute the topology of the network. More specifically, we aim to quantify the distribution of linking number, $Lk$ (defined as the number of times a minicircles wraps around one other) and the valence, $v$ (defined as the number of minicircles that are linked with any other one). 
To measure these, we perform molecular dynamics simulations steered by our AFM data, with the aim of obtaining XYZ coordinates of the DNA segments making up minicircles within the network.

To do this we select 1 $\mu$m x 1 $\mu$m region (ROI) within the kDNA cap (Fig.~\ref{fig:MD}A). We then binarise the image by  selecting the pixels whose intensity is larger than the mean background intensity plus 3 standard deviations. We then use this binarised image as a mask on the original ROI to extract the intensity of the pixels corresponding to DNA strands in the AFM image. These pixels are transformed into three types of phantom (non sterically interacting) and static (non moving) beads which attract the simulated DNA rings (see SI for details of the potentials used). 

We initialise a molecular dynamics simulation by placing $M$ perfectly circular minicircles within the 1 $\mu$m$^2$ region (see Fig.\ref{fig:MD}B). Each minicircle is modelled as a semi-flexible bead-spring polymer where each bead is $\sigma = 10$ nm (the AFM pixel resolution); in turn, the $2.5$ kb-long or $850$ nm-long minicircles are simulated with $85$ beads and a persistence length $l_p = 5\sigma = 50$ nm. The interaction between rings is modelled via a soft potential. 
Finally, the simulation is performed within a slab confinement in the $z$ direction with height $h=3.0 \sigma$. The steered simulation is split in 3 parts: (i) equilibration, (ii) steering and (iii) resolving crossings. In part (i) we homogenise the distribution of rings in the system. We thus set a low energy barrier for crossing and the rings are let to equilibrate. In part (ii) we steer the rings' coordinates by turning on attraction between the beads forming the DNA rings and the ``phantom'' beads obtained from the AFM image (see Fig.~\ref{fig:MD}B-C, see SI for details). This phase ensures that the simulated minicircle assume conformations that are compatible with the underlying AFM image. In part (iii) we resolve overlaps between rings by ramping up the height of the soft repulsive potential between polymer beads. The final output of this procedure is an ensemble of minicircle coordinates that do not display overlaps, with well-defined over/under-crossings, and whose 2D projection is compatible with the underlying AFM  (Fig.~\ref{fig:MD}C,D). 

Motivated by our previous findings, we perform the procedure just described in a region near the centre, and one at the periphery (rim excluded) of the kDNA. In the former, we initialise $M=90$ rings while at the periphery we initialise $M=80$ rings within the $1 \mu$m$^2$ ROI, in line with the values of minicircle density reported in Fig.~\ref{fig:panel_density}. To benchmark our steered simulations with experiments, we first compare the distribution of ring sizes. We find that the distribution is close to the one obtained from isolated kDNA minicircles in the AFM images (Fig.~\ref{fig:panel_single_rings}) with the caveat that the simulated rings display a broader size distribution and a smaller mean, which is reasonable given that they are in dense conditions whereas the AFM minicircles over which we compute the $R_g$ are isolated, outside the network. The experimental mean is $R_g = 83 \pm 4$ nm, while the simulations give $R_g=71 \pm 15$ at the centre and $R_g=72 \pm 15$ at the periphery (Fig.\ref{fig:MD}F). 

From the ensemble of conformations with resolved overlaps, we can unambiguously compute the Gauss linking number between pairs of minicircles as
\begin{equation}
Lk = \dfrac{1}{4 \pi} \oint_{\gamma_1} \oint_{\gamma_2} \dfrac{(\bm{r}_1 - \bm{r}_2) \cdot (d\bm{r}_1 \times d\bm{r}_2)}{|\bm{r}_1 - \bm{r}_2|^3}
\end{equation}
where $\bm{r}_i$ represents the 3D coordinate of curve $\gamma_i$. We thus define a linking matrix $Lk(i,j)$ where each entry is the number of times ring $i$ is linked to ring $j$. Additionally we define the valence of ring $i$ as $m_i = \sum_j \theta(|Lk(i,j)|)$ where $\theta(x)=1$ if $x>0$ and 0 otherwise. Interestingly, we find that the distribution of the valence depends, albeit weakly, on the distance from the centre of the network. In the more central ROI we find a mean valence $\langle m_{\rm cen} \rangle = 3.1 \pm 1.5$ while in the more peripheral region we find $\langle m_{\rm per} \rangle = 2.5 \pm 1.3$ Fig.~\ref{fig:MD}E). Further, we measure the distribution of linking number $lk=Lk(i,j)$ across all pairs and find that among those that are linked, the majority are singly linked, around 10\% doubly linked and less than 1\% triply linked.

These numbers are in good agreement with the bulk, indirect measures of Ref.~\cite{Chen1995} whereas here we can provide a single-molecule quantification. Importantly, our method does not assume {\em a priori} an ordered lattice arrangement of the rings. Indeed, we find that they form connected, percolating components with valence 3 even in absence of a precise hexagonal lattice structure. Instead, we find a broad distribution of valences which overall retain the percolating nature of the structure (Fig.~\ref{fig:MD}F). Our results thus confirm that the kDNA minicircles have on average valence 3, as found in Ref.~\cite{Chen1995}, but they also indicate a broad valence distribution and no crystalline order.

\subsection*{Elasticity of kDNA as a sub-isostatic network}

In light of our results, kDNAs can be thought of as a 2D elastic networks with nodes (the minicircles) that have valence around 3. In general, a network is said to be isostatic~\cite{Maxwell1864} when the number of constraints matches the number of degrees of freedom. For 2D networks, the critical isostatic coordination number is $v_c = 4$. Thus, the kDNA is a sub-isostatic (floppy) network, with valence comparable to that of other biological networks, such as collagen~\cite{Sharma2016}. The difference with collagen is that the bonds between nodes are not made by stiff fibres but are made by the linkages between minicircles, and their stiffness is approximately (at least for small strain) that of an entropic spring with constant $\kappa_0 = 3k_BT/R_g^2$.  Sub-isostatic networks display soft modes that cost zero energy even when weakly strained~\cite{Wyart2008} and undergo stiffening when stretched beyond a critical strain $\gamma_c(v)$~\cite{Sharma2016}. For strains $\gamma > \gamma_c$, we can estimate the bulk (area) stretch modulus as~\cite{Zaccone2011,Sharma2016} 
\begin{equation}
    Y = \dfrac{5}{48} \rho \kappa_0 R_g^2 |v-4| \simeq 0.1 \dfrac{pN}{\mu m} \, .
    \label{eq:stretchmodulus}
\end{equation}
In other words, we expect that it would take a modest force, around 3 pN, to stretch/compress a flat kDNA by 10\%. We note that this calculation does not account for the redundantly linked and denser rim around the network. The bending rigidity can then be approximated as $\kappa_{bend} \simeq Y (2 R_g)^2 \simeq 3 \, 10^{-21}$ J, being $2R_g$ about the average thickness of the kDNA. 

The bending rigidity $\kappa$ was also recently estimated using microfluidic constriction experiments and, in analogy with vescicles deformation, it was found to be $\kappa = 1.8 \, 10^{-19}$ N m~\cite{Klotz2020}. However, the combination of in-plane and out-of-plane deformations in kDNA is expected to be different from that of vescicles and, more importantly, we expect drastically different area stretch modulus. Lipid bilayers are liquid-like compositions of small molecules with $\sim$ nm thickness and display large stretch moduli, $Y \simeq 0.1-1$ N/m~\cite{Janshoff2015} and equally large bending stiffnesses $\kappa \simeq 10^{-18}$ N m. 
In contrast, kDNA is made of 2.5 kbp-long DNA rings with an average size $R_g \simeq 85$ nm and the density of material inside the kDNA is low compared with lipid bilayers, rendering the structure much easier to deform both in-plane and out-of-plane. for these reasons, we expect the stretch modulus and bending stiffness of kDNA to be widely different from that of vescicles. According to Eq.~\eqref{eq:stretchmodulus} we expect these to be two/three orders of magnitude smaller than lipid vescicles.
A clear complication, that yields an apparent long-time stability of kDNA shapes (and thus apprently large bending rigidity) is that kDNA networks in the bulk have undergone a buckling transition due to the border constriction. 
In fact, the autocorrelation of kDNA anisotropy shows fast, sub-second rearrangements which are more in line with far smaller and more flexible molecules~\cite{Klotz2020}.  

Further, we note that the buckling behaviour of a 2D elastic thermal sheet is typically controlled by the dimensionless F\"{o}ppl-von K\'{a}rm\'{a}n number $v_K = A/h^2$, where $A$ is the area of the sheet and $h$ its height. The parameter $v_K$ can also be expressed in terms of the Young modulus and bending rigidity as $v_K = Y R^2/\kappa$, where $R$ is a characteristic linear dimension of the system. 
Taking $h$ to be the diameter of minicircles in their relaxed state {\it in vitro}, we obtain for kDNA networks $v_K \simeq 1700$ which is far lower than other 2D materials (for instance graphene has $10^9$, being extremely thin). 
In this respect, kDNA is considered to be ``thick'' and therefore easily stretchable/compressible before buckling. In the ``thin'' limit, buckling occurs before any in-plane deformation. 

The competition of compression and bending moduli gives rise to a natural lengthscale called ``thermal lengthscale'' which dictates the behaviour of 2D elastic thermal sheets~\cite{Shankar2021,Chen2022}. This lengthscale is found as
\begin{equation}
    l_{th} = \sqrt{\dfrac{16 \pi^3 \kappa^2}{3 k_B T Y}} \, .
    \label{eq:thermalscale}
\end{equation}
When compression deformations are larger than  $l_{th}$, it is more energetically favourable for a 2D elastic sheet to buckle. By using the values of $Y$ and $\kappa$ found above we obtain a thermal length-scale $l_{th} \simeq 1.8$ $\mu$m. This value may be interpreted as the amount of compression needed for the network to buckle. Interestingly, the ratio of maximum radius of the kDNA to this thermal lengthscale is $R_{g,max}/l_{th}=2.5$, which is in the buckled phase (see Fig.~\ref{fig:border_effect_sim}C). This implies that the buckling behaviour of kDNA is well described by the physics of 2D elastic thermal sheets and that, as we discussed above, the linking properties of the nodes at the rim are such that their relaxed state induces an in-plane compression beyond the thermal lengthscale of the network $l_{th}$, thereby inducing buckling.

\section*{Conclusions}
Overall, our study is the first to perform a quantitative analysis of single-molecule data on the structure and topology of \emph{C. Fasciculata} kinetopalst DNA networks. While previous works used indirect methods to obtain the kDNA topology~\cite{Chen1995} a single-molecule characterisation of kDNA structure and topology did not exist.

We have employed high-resolution AFM, quantitative image analysis and MD simulations to discover that the kDNA does not display a uniform DNA density but instead it has more minicircles in the middle of the network than the periphery (Fig.~\ref{fig:panel_density}). On average, we find about 95 minicircles per $\mu$m$^2$ in the cap of the network and 140 minicircles per $\mu$m$^2$ at the rim. Additionally, we have used morphological segmentation to quantify the pore size of the network (Fig.~\ref{fig:poresize}) and found that the mesh size is smaller in the middle (about 30 nm) compared with the periphery (about 36 nm). 

By noticing that the minicircles at the nodes appear stretched under AFM (also seen in previous EM~\cite{barker1980ultrastructure} and AFM~\cite{Cavalcanti2011} images), we argued that when not adsorbed onto a surface, the rim should shrink by $\simeq 2$ fold, due to the entropic elasticity of the minicircles. Motivated by this, we simulated the behaviour of a chainmail of linked rigid rings under varying degrees of constraint on the size $R_{g,rim}$ of the border, and observed a buckling transition when $R_{g,rim}$ was set to be around 2.5-3 times smaller than that of the fully flat kDNA (Fig.~\ref{fig:border_effect_sim}). The buckling transition seen around $R_{g,max}/R_{g,rim} \gtrsim 2$ is in good agreement with the expected entropic shrinking of the kDNA in bulk and thus explains the stable ``shower cap'' buckled shape recently seen in confocal microscopy~\cite{Klotz2020,Soh2020}. 
Both our experiments and simulations agree with the calculation of the thermal length-scale for kDNA being (Eq.~\eqref{eq:thermalscale}) $l_{th} \simeq 1.8$ $\mu$m; this is around $2.5$ times smaller than the radius of fully flat kDNA and it marks the transition where buckling (out-of-plane) deformations are favoured over compression (in-plane) deformations.  

Finally, we have used steered molecular dynamics simulations to obtain ensembles of rings conformations that are compatible with the DNA distributions in the AFM images and can resolve certain topological ambiguities that cannot be resolved in the AFM image. Using these simulations we have independently measured the valence of the minicircles in the network and found that it displays a broad distribution with mean around 3. This finding is in remarkable good agreement with the measures by Cozzarelli~\cite{Chen1995} in spite of the fact that they are obtained in two completely different methods. Differently from the indirect, bulk quantification of the network topology done in the past, our high-resolution quantitative imaging allowed us to discover that the the topology and connectivity of the network (i) is heterogeneous and broadly distributed and (ii) depends on the distance from the centre of the network. It would be interesting in the future to understand more about the mechanisms leading to this gradient. Notably, our high-resolution and MD approach yielded networks that do not resemble perfect hexagonal arrangements but are instead random (Fig.~\ref{fig:MD}). In the work of Cozzarelli~\cite{Chen1995}, the hexagonal arrangement model was imposed due to the assumption of a perfectly two dimensional network. We argue that this approximation is too stringent, and that the percolating nature of the kDNA can be achieved also allowing rings to randomly link at the right density~\cite{Diao2012,Michieletto2014a}. 

We note that in the language of 2D random networks, a valence (or coordination number) $v=3$ is below the isostatic value~\cite{Maxwell1864}, that for 2D networks is $v_c=4$. This renders the kDNA a sub-isostatic, floppy network with soft (zero energy) modes and zero stress response at strains $\gamma$ below a critical $\gamma_c(v)$~\cite{Wyart2008}. At the same time, and although kDNA networks may resemble suspended membranes or lipid bilayers, they display a highly unusual structure, made of DNA minicircles that are thousands of base-pairs long. More specifically, compared with lipid bilayers, kDNA displays a lower density and larger thickness. For this reason, we expect its material properties to be markedly different from that of lipid membranes, which are essentially incompressible~\cite{Janshoff2015}.
Indeed, we estimated the kDNA stretch modulus to be around (Eq.~\eqref{eq:stretchmodulus}) $Y \simeq 0.1$ pN/$\mu$m and its bending stiffness $\kappa \simeq 1$ pN nm, both thousands of times smaller than those of lipid membranes.    

The evidence suggesting that the minicircles in the kDNA have valence around 3 is intriguing. A random network with valence 3 is poised near the critical percolation point~\cite{Michieletto2014a,Diao2012b,Diao2015} yet below the isostatic point for the onset of rigidity~\cite{Maxwell1864}. Being poised closed to the percolation point ensures that the network is overall connected (thus preserving the integrity of the genome during replication) yet avoids the generation of redundant constraints or a topologically frustrated ``over-linked'' and rigid network~\cite{Michieletto2014a}. 
Perhaps even more intriguingly, the volume fraction of kDNA \emph{in vivo} suggests an overlap number $P \simeq 10$, in turn suggesting that kDNA minicircles should display a far larger valence if simply allowed to cross each other freely. This argument suggests that the topology of the network is controlled {\it in vivo}. In this respect, packaging proteins such as KAP and controlling Topoisomerase activity may play a key role~\cite{Yaffe2021}. 

We also mention that although different species of Trypanosomes have different kDNA structures, they all display an overall percolating network. We argue that species with longer minicircles should display an even larger valence, scaling as $v \sim \rho L^{3\nu-1}$~\cite{Cates1986,Vilgis1997} with $\nu=1/2$ for short rings and $\nu=1/3$ for longer flexible rings~\cite{Halverson2011statics}. If this were not to be the case, it would be a strong evidence for a biological control of kDNA topology implying an evolutionary benefit in keeping $v \simeq 3$.    

In summary, we have here reported here a single-molecule high-resolution quantitative analysis of one of the most fascinating genomes in nature. We hope that our work will not only help to unveil the self-assembly and topological regulation of generic Kinetoplast DNA networks and their evolutionary pathway but also provide some insights on how to synthetically design 2D topological soft materials.  

\section*{Methods}

In order to obtain high-resolution information on the kDNA structure we perform Atomic Force Microscopy (AFM) on kDNA samples purified from \emph{C. fasciculata} (Inspiralis). The kDNA sample was diluted to a concentration of 50-100 ng/$\mu$L in a buffer solution containing 50 mM MgCl$_2$ then a droplet of it was deposited onto the mica surface for 1 min followed by 1 mL deionized water flushing and nitrogen blowing. Imaging was performed on a Bruker Multimode AFM in Peakforce-HR mode, using Bruker Scanasyst-air-HR cantilevers with a nominal resonant frequency of 130 kHz and  spring constant of 0.4 N/m. 

In AFM images, the intensity of the pixel is a direct measure of its height: brighter pixels correspond to crossings and overlaps of DNA strands. Although the the apparent DNA height and width is affected by the tip force, tip radius and non-hydrated conditions, we use isolated plasmids - similarly  affected by artefacts - as a volume reference. Thanks to this feature, we can directly map height to DNA density in each pixel. There can be cases in which DNA strands (about 2-5 nm wide depending on salt conditions) lay side-by-side in a 10 nm pixel. In these cases the intensity of the pixel is not directly proportional to the underlying mass of DNA. The reference volume is measured on isolated plasmids that are much less likely to have multiple strands in one pixel. Therefore, we expect to slightly underestimate the true DNA density in the network.

\subsection{Morphological segmentation} 
Used MorphoJLib with no noise reduction and tolerance 15. This plugin uses a modified watershed algorithm to identify objects as basins (the pores) separated by boundaries (the DNA strands). An image with overlaid basins was then generated (see Fig.~\ref{fig:poresize}A-B) and analysed with the ``analyse region'' function of MorphoLibJ which returns a list of the values of area, perimeter, circularity and centre of mass of all the basins. The values of pore sizes were then obtained by taking the square root of the areas. The artefacts inside the countour of the kDNA were then removed by identifying the outliers with very large area.

\subsection{Simulations with border constriction}
The networks are first built using the NetworkX~\cite{hagberg2008exploring} Python package, and the corresponding meshes are analyzed using  libIGL~\cite{fang2022topocut} for Python. The networks are built starting from a planar configuration by placing on half the nodes of the network, corresponding to second neighbours, planar rings and the joining those through the random placement of a set of distorted rings on the remaining nodes. This procedure ensures that the sign of the Hopf link between any two rings is picked randomly between $-1$ and $+1$, avoiding the onset of topological phenomena such as those reported in ~\cite{tubiana2021ring}. Using this strategy we produced three different topologies. The border is then identified with the set of rings that are linked only to two more rings, or that are directly linked to one such ring. More details are reported in the SI. The kDNA minicircles are then modelled as semi-rigid Kremer-Grest polymers~\cite{Kremer1990} made of $m=60$ beads having diameter $\sigma$ and connected by FENE bonds. The rings have a persistence length $l_p = 120\sigma$. Each network is a circular hexagonal patch composed by $n=604$ rings. Different rings interact only by excluded volume, modelled through a WCA potential. The system is evolved using an underdamped Langevin Dynamics with timestep $dt=0.01\tau_{LJ}$ and damping $\gamma=0.1\tau_{LJ}^{-1}$, where $\tau_{LJ}=$ is the characteristic time of the simulation. At each timestep, we impose the constraint potential $V_{constr}= K(R_g-R_{g,hc})^2$. These simulations are performed in LAMMPS~\cite{Plimpton1995}. the codes can be found ope source at \url{https://github.com/luca-tubiana/kDNA-border-sims}.

\subsection{AFM-steered simulations} 
Briefly, we model kDNA minicircles as bead-spring polymers with a persistence length of $50$ nm. Each bead is given a size equal to that of the resolution of the pixel, i.e. $\sigma=10$ nm. The AFM image is transformed (see main text and SI) into a series of phantom, static beads that act as attractors of the DNA beads. The system is evolved using a velocity-Verlet algorithm and Langevin dynamics (implicit solvent) with timestep $dt=0.01 \tau_{Br}$, where $\tau_{Br} = \gamma \sigma^2/k_BT$ is the Brownian time. For more details on the force fields used, see SI. These simulations are also performed in LAMMPS. The codes can be found open source at \url{https://git.ecdf.ed.ac.uk/taplab/kdna-afm-md}.

\section*{Acknowledgements}
This project has received funding from the European Research Council (ERC) under the European Union's Horizon 2020 research and innovation programme (grant agreement No 947918, TAP). DM also acknowledges support of the Royal Society via a University Research Fellowship. LT acknowledges support from MIUR, Rita Levi Montalcini Grant, 2016. CD acknowledges support of ERC Advanced Grant no. 883684. PH would like to acknowledge Prof. Yunfei Chen, Prof. Zhonghua Ni and support from the China Scholarship Council (CSC201906090029). The authors acknowledge insightful discussions with Jaco van der Torre, Alice Pyne and Raffaello Potestio. The authors also acknowledge the contribution of the COST Action Eutopia, CA17139. 

\bibliography{library}

\end{document}